\documentclass{article}

\usepackage{arxiv}

\usepackage[utf8]{inputenc} 
\usepackage[T1]{fontenc}    
\usepackage{hyperref}       
\usepackage{url}            
\usepackage{booktabs}       
\usepackage{amsfonts}       
\usepackage{nicefrac}       
\usepackage{microtype}      
\usepackage{lipsum}		
\usepackage{graphicx}
\usepackage{doi}
\usepackage{float}
\usepackage{geometry}
\usepackage{longtable}

\title{2024 Google Scholar Research Interest Ranking for Top 3260 Computer Science Authors}

\author{ 
    \href{https://orcid.org/0009-0008-0056-0557}{\includegraphics[scale=0.06]{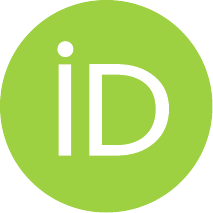}\hspace{1mm}Atharva Rasane}\thanks{Department of Computer Science and Engineering, KLE Technological University’s, Dr. M. S. Sheshgiri College, Belagavi, India. Email: \texttt{rratharva@gmail.com}} \\
    Dept. Computer Science and Engineering\\
    KLE Technological University’s,\\
    Dr. M. S. Sheshgiri College\\
    Belagavi, India \\
    \texttt{rratharva@gmail.com} \\
}

\renewcommand{\headeright}{}
\renewcommand{\undertitle}{}
\renewcommand{\shorttitle}{Google Scholar Research Interest Ranking for Computer Science}

\hypersetup{
pdftitle={A template for the arxiv style},
pdfsubject={q-bio.NC, q-bio.QM},
pdfauthor={David S.~Hippocampus, Elias D.~Striatum},
pdfkeywords={First keyword, Second keyword, More},
}

\begin{document}
\maketitle

\begin{abstract}
	Computer science research has grown increasingly diverse, with authors investigating a broad spectrum of topics. This paper analyzes the self-reported research interests of the top 3,260 most cited computer science scholars, according to Google Scholar. By leveraging the \texttt{scholarly} Python library, we systematically retrieved and classified each researcher’s interests into predefined categories grounded in the Computer Science Ontology (CSO). Our results reveal a clear hierarchy of primary focus areas such as Artificial Intelligence, Software Engineering, Data Mining, and Computer Systems. We further examine how these interests are distributed, shedding light on which fields are emerging as popular, which are well-established, and which receive comparatively less attention. These findings provide an up-to-date snapshot of key research trends and lay the groundwork for guiding future studies in computer science.
\end{abstract}


\section{Introduction}
Computer science is the study of the theoretical foundations, experimental processes, and engineering principles that underpin the design and use of computers \cite{wikiCompSci2024}. This field encompasses many sub-disciplines, including artificial intelligence, software design, and computer networks. By examining research trends, we can gain valuable insights into the future trajectory of computer science. One effective way to identify these trends is to analyze the self-reported research interests of leading experts. In this paper, we leverage the \texttt{scholarly} Python package to gather the research interests of the 3,260 most cited authors in computer science, according to Google Scholar. Through this approach, we aim to determine which areas attract the highest level of attention, which are emerging as popular or in-demand topics, and which receive comparatively less focus. Ultimately, our goal is twofold: to provide an overview of the current state of computer science research and to outline a methodology for collecting and analyzing research fields in a systematic and scalable manner.

\section{Literature Review}

In the rapidly evolving field of research tools and ontologies, several notable contributions have reshaped the way researchers conduct and present their work. The \texttt{scholarly} Python package \cite{scholarly} stands out as a convenient tool for retrieving Google Scholar data, including citations, author profiles, and publications. This functionality greatly simplifies academic research workflows by automating tasks that would otherwise be time consuming when done manually. 

Another significant development is the Computer Science Ontology (CSO) \cite{cso}, which organizes the vast domain of computer science by applying the Klink-2 algorithm to a large corpus of scientific articles. CSO provides a detailed, automatically generated taxonomy with over 14,000 topics and 162,000 relationships, enabling researchers to better explore, categorize, and understand various subfields within computer science.

For tasks involving natural language processing and text embeddings, the SentenceTransformers library \cite{sentence-transformers} has proven to be indispensable. It creates dense vector representations of sentences and paragraphs, facilitating applications like semantic similarity, information retrieval, and clustering. Within SentenceTransformers, the \texttt{all-MiniLM-L6-v2} model \cite{all-MiniLM-L6-v2} strikes an excellent balance between speed and accuracy, making it a practical choice for situations where computational resources are limited yet reliable performance is crucial.

In the realm of data manipulation and visualization, Pandas \cite{pandas} remains a premier solution for managing structured data. Its user-friendly DataFrame structure and extensive suite of functions allow researchers to clean, process, and analyze datasets with relative ease. Complementing these capabilities, Matplotlib \cite{matplotlib} provides an extensive toolkit for creating static, interactive, and animated visualizations. Its customizable plots facilitate clear and compelling presentation of research findings.

Together, these tools enhance the research process by simplifying data acquisition, refining data exploration, and enabling insightful presentation of results. They collectively empower researchers to devote more time to innovation and discovery, driving progress across a broad spectrum of computational and scientific endeavors.

\section{Methodology}
We first collected the data by using the Python library \texttt{scholarly}; we chose the search query as ``computer science'' and used the \texttt{search\_author()} function of \texttt{scholarly} to get an iterator over the author profiles that match the search query. We iterated over all the authors until an exception occurred, which happened at author number 3261. (Note: Authors were iterated from the most citations to the least and thus we obtained the data of the top 3260 authors in computer science.) The data structure returned by \texttt{scholarly} was a dictionary with the following keys:

\begin{itemize}
    \item \textbf{container\_type}: Specifies the type of data container. In this case, it's ``Author'', indicating that the data pertains to an author profile.
    \item \textbf{filled}: A list that tracks which sections of the author's profile have been populated with detailed information. In this case, an empty list (\texttt{[]}) suggests that only essential information is present and that additional details can be fetched using the \texttt{fill} method.
    \item \textbf{source}: Denotes the origin of the data. Here, it's set to \texttt{<AuthorSource.SEARCH\_AUTHOR\_SNIPPETS: 'SEARCH\_AUTHOR\_SNIPPETS'>}, indicating that the information was obtained from a search result snippet.
    \item \textbf{scholar\_id}: A unique identifier assigned to the author by Google Scholar. This ID can be used to retrieve more detailed information about the author.
    \item \textbf{url\_picture}: The URL of the author's profile picture on Google Scholar.
    \item \textbf{name}: The author's full name.
    \item \textbf{affiliation}: The author's current institutional affiliation (``Fellow, Joint Center for Quantum Information and Computer Science (QuICS), NIST/UMD'').
    \item \textbf{email\_domain}: The domain of the author's email address.
    \item \textbf{interests}: A list of the author's research interests.
    \item \textbf{citedby}: The total number of citations the author has received, reflecting their impact on the academic community.
\end{itemize}

All the authors were then appended to a list and converted to a Pandas data frame using the \texttt{DataFrame()} function. They were then stored in a CSV file for further processing.

Here, we face our first problem: Google Scholar doesn't have a fixed research instrument and is instead typed in by the authors. To address this issue, we must classify the interests into fixed ones. To do this, we first need to define fixed interest categories. To achieve this, we refer to the \textbf{Computer Science Ontology (CSO)}, which is an automatically generated taxonomy of research topics in the field of computer science produced by Open University in collaboration with Springer Nature. We first classify the research topics into a list of categories consisting of first-degree relations and second-degree relations of computer science research topics as per CSO 3.4. To classify into these topics, we utilize the \texttt{SentenceTransformer} library, which is built on top of HuggingFace's transformers and specializes in generating embeddings for sentences or phrases. The model used here is \texttt{all-MiniLM-L6-v2}, which is pre-trained on a variety of natural language tasks using datasets that involve textual similarity, making it well-suited for comparing sentences or phrases. We use it to generate encodings for categories and interests. Once we encode the categories and interests into embeddings, we compute cosine similarity between the interests and all the categories using \texttt{util.cos\_sim}, and we categorize into the appropriate category. Finally, we count all the categories and visualize them as bar charts and word clouds to find the top research fields. The categories into which we are classified are as follows:

\begin{itemize}
    \item \textbf{artificial\_intelligence}: The simulation of human intelligence processes by machines, especially computer systems, enabling them to perform tasks such as learning, reasoning, and problem-solving.
    \item \textbf{robotics}: An interdisciplinary branch that integrates computer science and engineering to design, construct, operate, and use robots, aiming to automate tasks in various environments.
    \item \textbf{hardware}: The physical components of a computer system, including the central processing unit (CPU), memory, storage devices, and peripherals like keyboards and monitors.
    \item \textbf{computer\_operating\_systems}: Software that manages computer hardware and software resources, providing standard services for computer programs and acting as an intermediary between users and the computer hardware.
    \item \textbf{computer\_networks}: Interconnected systems that allow computers to communicate and share resources, data, and applications, facilitating functions like file sharing and internet access.
    \item \textbf{bioinformatics}: An interdisciplinary field that develops methods and software tools for understanding biological data, combining biology, computer science, and information technology to analyze and interpret biological information.
    \item \textbf{software\_engineering}: The systematic application of engineering approaches to the development, operation, and maintenance of software, ensuring it is reliable, efficient, and meets user requirements.
    \item \textbf{information\_technology}: The use of computers, storage, networking devices, and other physical devices, infrastructure, and processes to create, process, store, secure, and exchange electronic data.
    \item \textbf{data\_mining}: The process of discovering patterns, correlations, and anomalies within large data sets to predict outcomes using methods at the intersection of machine learning, statistics, and database systems.
    \item \textbf{information\_retrieval}: The process of obtaining relevant information from large repositories, such as databases or the internet, based on user queries, enabling efficient access to needed data.
    \item \textbf{computer\_programming}: The process of designing and building executable computer programs to accomplish specific computing tasks, involving tasks such as analysis, generating algorithms, and implementing them in programming languages.
    \item \textbf{human\_computer\_interaction}: The study and practice of designing, implementing, and evaluating interactive computing systems for human use, focusing on the interfaces between people and computers.
    \item \textbf{computer\_aided\_design}: The use of computer systems to assist in the creation, modification, analysis, or optimization of a design, improving the productivity of designers and the quality of designs.
    \item \textbf{computer\_security}: The protection of computer systems and networks from information disclosure, theft of or damage to their hardware, software, or data, and disruption or misdirection of their services.
    \item \textbf{graph\_theory}: A branch of mathematics and computer science concerned with the properties of graphs, which are structures made up of vertices connected by edges, used to model pairwise relations between objects.
    \item \textbf{theoretical\_computer\_science}: A division of computer science that focuses on the abstract and mathematical aspects of computing, including algorithms, computation theory, and the limits of what can be computed.
    \item \textbf{computer\_system}: An individual computing device that combines hardware, software, and peripheral devices to perform computing tasks.
    \item \textbf{computer\_networks}: Integrated systems that combine hardware, software, and peripheral devices to perform computing tasks, encompassing personal computers to large-scale enterprise servers.
    \item \textbf{internet}: A global network of interconnected computers that communicate through standardized protocols, enabling the exchange of data and access to information and services worldwide.
    \item \textbf{computer\_hardware}: The tangible, physical components that make up a computer system, such as the motherboard, CPU, RAM, hard drives, and input/output devices.
    \item \textbf{software}: A collection of data or computer instructions that tell the computer how to work, including applications, scripts, and programs that run on a device.
    \item \textbf{computational\_time}: The amount of time a computer takes to execute an algorithm is often analyzed in terms of time complexity to determine efficiency.
    \item \textbf{computational\_costs}: The resources required to execute a computational process, including time, memory, and energy, influence the feasibility and efficiency of algorithms.
    \item \textbf{computation\_time}: The computer's duration to perform a given task or solve a problem, synonymous with computational time.
    \item \textbf{computational\_efficiency}: A measure of the resource usage of an algorithm or process, aiming to minimize computational time and memory usage while maximizing performance.
    \item \textbf{computer\_network}: A digital telecommunications network that allows computers to exchange data and share resources, facilitating communication between interconnected devices.
    \item \textbf{computer\_communication\_networks}: Systems that enable data communication between different computing devices over various transmission media, facilitating resource sharing and information exchange.
    \item \textbf{operating\_system}: Software that manages computer hardware and software resources, provides services for computer programs, and acts as an intermediary between users and the computer hardware.
    \item \textbf{computation\_efficiency}: A measure of how effectively computational resources are utilized to achieve a desired outcome, focusing on optimizing performance while minimizing resource consumption.
    \item \textbf{formal\_languages\_and\_automata\_theory}: A branch of computer science that deals with the definitions and properties of formal languages (sets of strings) and the computational machines (automata) that recognize them, fundamental to compiler design and parsing.
    \item \textbf{medical\_informatics}: The intersection of information science, computer science, and healthcare, focusing on the resources, devices, and methods required to optimize the acquisition, storage, retrieval, and use of information in health and biomedicine.
    \item \textbf{computer\_imaging\_and\_vision}: Fields that involve processing and interpreting visual information from the real world through computers, enabling tasks such as image recognition, analysis, and reconstruction.
    \item \textbf{pattern\_matching}: The act of checking a given sequence of tokens for the presence of the constituents of some pattern, used in algorithms for searching and manipulating text and data.
\end{itemize}

\section{Results}

\begin{figure}[H]
	\centering
	\includegraphics[width=\textwidth]{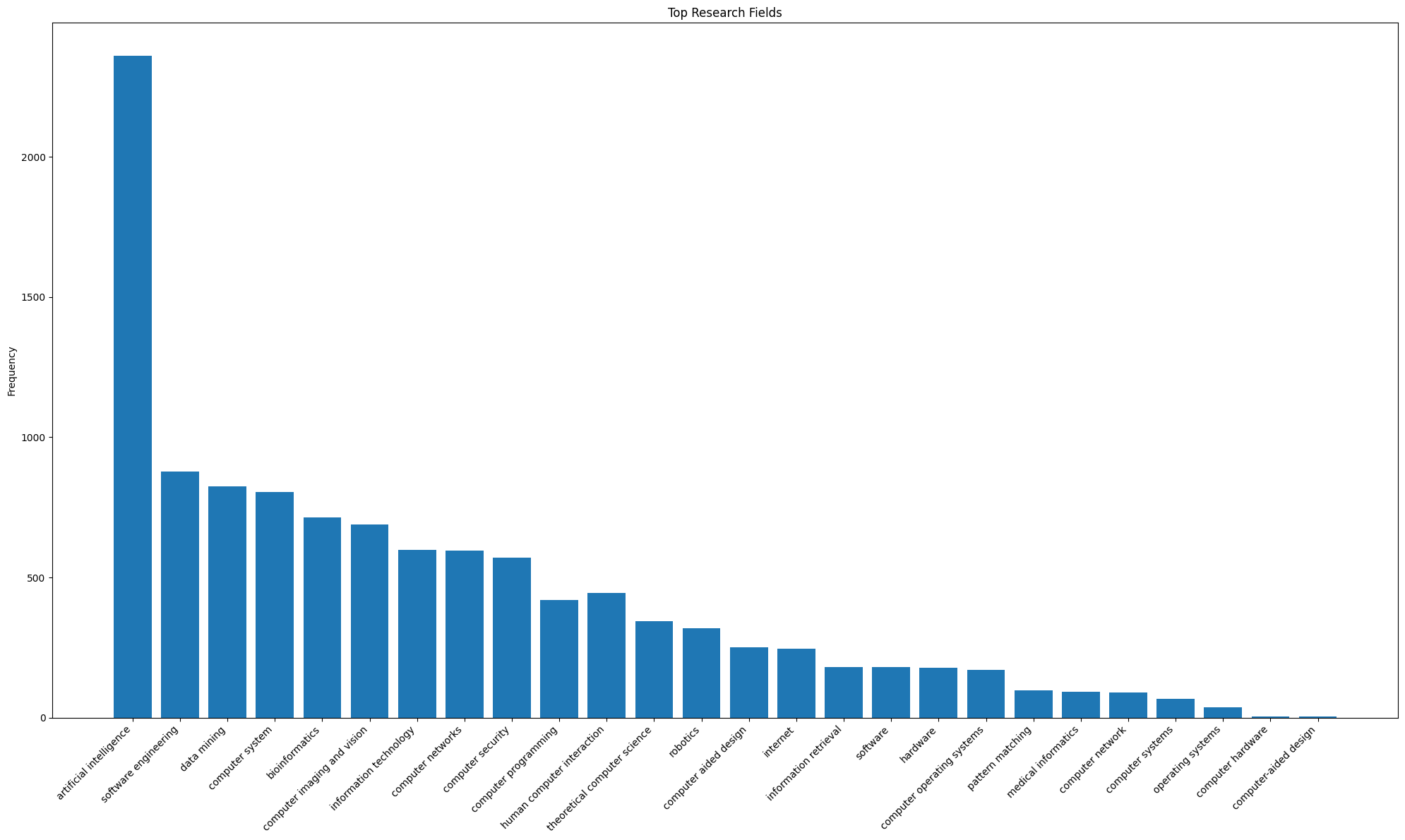}
	\caption{Top Computer Science Research Fields}
	\label{fig:fig1}
\end{figure}

\begin{figure}[H]
	\centering
	\includegraphics[width=\textwidth]{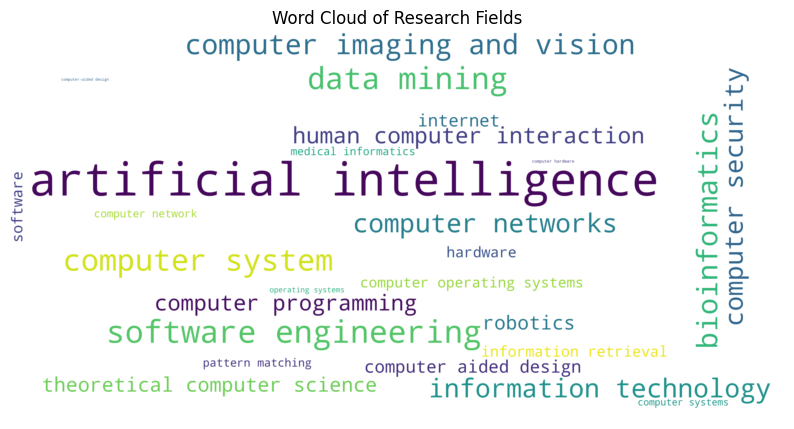}
	\caption{Word Cloud of Top Computer Science Research Fields}
	\label{fig:fig2}
\end{figure}

\begin{longtable}{|l|r|r|}
\hline
\textbf{Topic} & \textbf{Count} & \textbf{Percentage (\%)} \\ \hline
artificial intelligence & 2360 & 21.14 \\ \hline
software engineering & 877 & 7.86 \\ \hline
data mining & 826 & 7.40 \\ \hline
computer system & 805 & 7.21 \\ \hline
bioinformatics & 714 & 6.40 \\ \hline
computer imaging and vision & 688 & 6.16 \\ \hline
information technology & 598 & 5.36 \\ \hline
computer networks & 597 & 5.35 \\ \hline
computer security & 572 & 5.12 \\ \hline
computer programming & 419 & 3.75 \\ \hline
human computer interaction & 445 & 3.98 \\ \hline
theoretical computer science & 343 & 3.07 \\ \hline
robotics & 319 & 2.86 \\ \hline
computer aided design & 248 & 2.22 \\ \hline
internet & 247 & 2.21 \\ \hline
information retrieval & 181 & 1.62 \\ \hline
software & 181 & 1.62 \\ \hline
hardware & 177 & 1.59 \\ \hline
computer operating systems & 170 & 1.52 \\ \hline
pattern matching & 98 & 0.88 \\ \hline
medical informatics & 92 & 0.82 \\ \hline
computer network & 89 & 0.80 \\ \hline
computer systems & 68 & 0.61 \\ \hline
operating systems & 38 & 0.34 \\ \hline
computer hardware & 5 & 0.04 \\ \hline
computer-aided design & 4 & 0.03 \\ \hline
\end{longtable}

The 2024 Google Scholar Research Interest Ranking for Computer Science reveals a clear hierarchy of prevalent research areas within the field. \textbf{Artificial Intelligence} dominates the landscape, accounting for 21.14\% of the total interests, highlighting its significant influence and the widespread focus on machine learning, reasoning, and intelligent system development. Following closely are \textbf{Software Engineering} (7.86\%), \textbf{Data Mining} (7.40\%), and \textbf{Computer Systems} (7.21\%), each representing substantial portions of research activity and underscoring the importance of developing robust software methodologies, extracting meaningful patterns from large datasets, and enhancing the foundational infrastructure of computing environments. Other notable areas include \textbf{Bioinformatics} (6.40\%) and \textbf{Computer Imaging and Vision} (6.16\%), reflecting the interdisciplinary nature of modern computer science and its applications in biology and visual data processing. Additionally, fields such as \textbf{Information Technology}, \textbf{Computer Networks}, and \textbf{Computer Security} each hold between 5\% and 6\% of the interests, indicating ongoing efforts to advance connectivity, protect digital assets, and manage information effectively. The ranking also highlights emerging and specialized topics like \textbf{Pattern Matching} (0.88\%) and \textbf{Medical Informatics} (0.82\%), which, although smaller in scale, contribute to the diversification and expansion of computer science research. Overall, the distribution of research interests emphasizes the dominance of artificial intelligence while also showcasing a balanced emphasis on both foundational and applied areas, reflecting the dynamic and multifaceted nature of the computer science discipline.

\section{Conclusion}
Our analysis offers a systematic overview of the current landscape in computer science research by examining the self-reported interests of highly cited scholars. The prominence of Artificial Intelligence, Software Engineering, and Data Mining underscores the community's sustained focus on cutting-edge computational methods, large-scale data analysis, and secure, efficient systems. At the same time, emergent topics like Pattern Matching and Medical Informatics illustrate ongoing diversification in the field, pointing to interdisciplinary ventures that continue to broaden the scope of computer science. By applying a structured approach grounded in Computer Science Ontology and facilitated by tools such as \texttt{scholarly} and SentenceTransformers, we have demonstrated a scalable methodology for collecting, categorizing, and visualizing research interests. Ultimately, this study highlights prevailing research directions and sets a foundation for future analysis, enabling researchers, institutions, and funding bodies to make more informed decisions about where to concentrate their efforts and resources.

\bibliographystyle{plain}






\end{document}